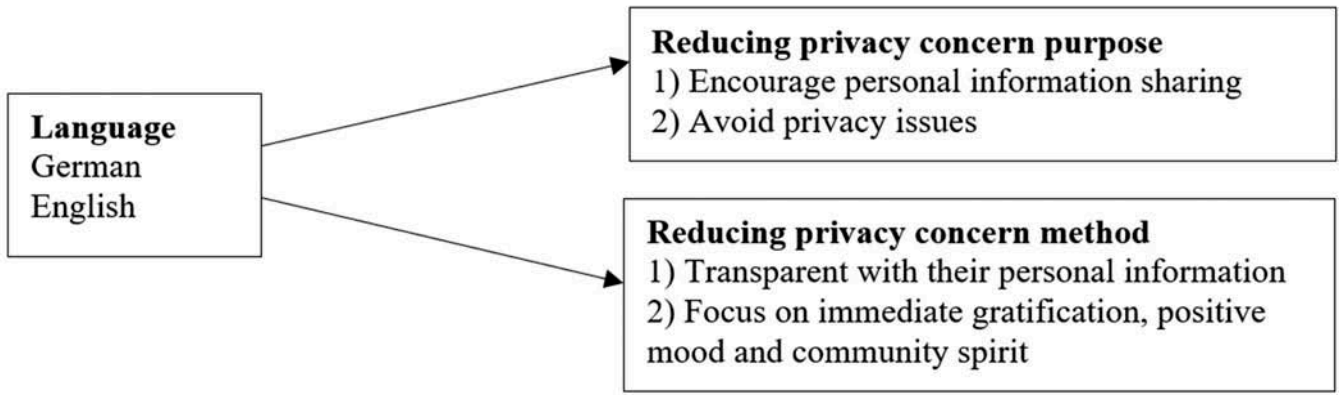

Fig. 1. The role of language in reducing privacy concern in Airbnb

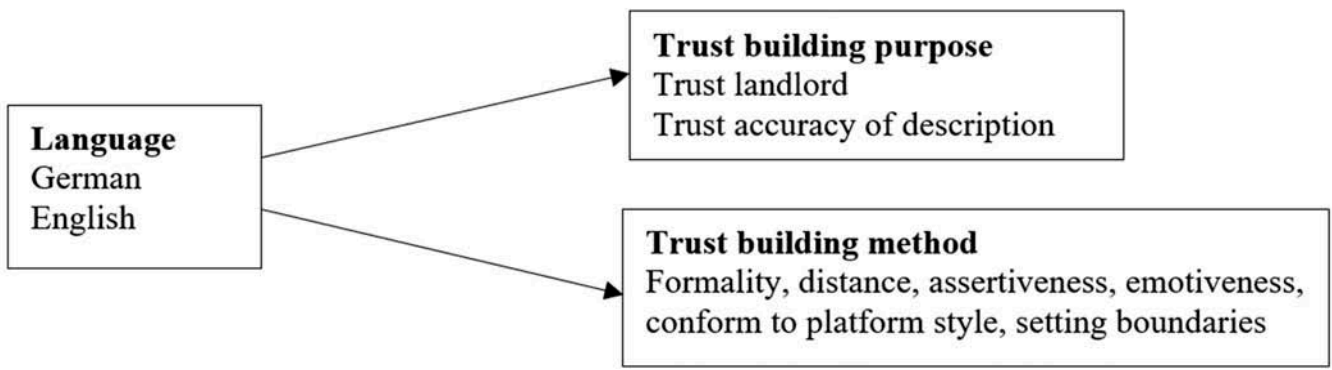

Fig. 2. The role of language in building trust in Airbnb

**OPERATIONS, INFORMATION & TECHNOLOGY | RESEARCH ARTICLE**

# Exploring the language of the sharing economy: Building trust and reducing privacy concern on Airbnb in German and English

Alex Zarifis, Richard Ingham and Julia Kroenung

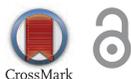

OPERATIONS, INFORMATION & TECHNOLOGY | RESEARCH ARTICLE

# Exploring the language of the sharing economy: Building trust and reducing privacy concern on Airbnb in German and English

Alex Zarifis[1]*, Richard Ingham[2] and Julia Kroenung[3]

**Abstract:** Several countries' economies have been disrupted by the sharing economy. Global champions like Airbnb and Uber use similar models and platforms across many countries. However, each country and its consumers have different characteristics including the language used. The text in the profile of those offering their properties in England in English and in Germany in German, are compared to explore whether trust is built, and privacy concerns are reduced in the same way. Six methods of building trust are used by the landlords: (1) the level of formality, (2) distance and proximity, (3) emotiveness and humor, (4) being assertive and passive aggressive, (5) conformity to the platform language style and terminology and (6) setting boundaries. Privacy concerns are not usually reduced directly as this is left to the platform. The findings indicate that language has a limited influence and the platform norms and habits are the biggest influence.

Subjects: Hospitality Marketing; Services Marketing; Marketing Research; International Marketing; Internet / Digital Marketing / e-Marketing; Marketing Communications; Marketing Management; Relationship Marketing; Retail Marketing

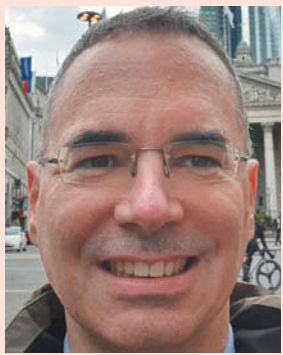

Alex Zarifis

ABOUT THE AUTHORS

Dr Alex Zarifis has taught and carried out research at several universities including the University of Manchester and the University of Mannheim. He obtained his PhD from the University of Manchester. His research interests are on e-business particularly trust and information privacy. His research has featured in journals such as Computers in Human Behavior and Information Technology & People.

Professor Richard Ingham, is a visiting professor at the University of Westminster. He holds degrees from the University of Oxford and the University of London and has taught at the University of Reading, Mannheim and Birmingham City University. He has published several books on language.

Dr Julia Kroenung is the Chair for E-Business and E-Government at the University of Mannheim. She completed her PhD at the Goethe University, Frankfurt in Germany. Her interests include e-business and social issues in IT. Her research has featured in journals such as Information & Management.

PUBLIC INTEREST STATEMENT

## 1. Introduction

The expansion of the sharing economy (SE), also referred to as the peer economy or collaborative consumption, is spreading around the world. SE is growing in popularity despite several challenges such as insufficient regulation or legal restrictions, and negative events such as fatal accidents and crimes against consumers. The push of SE champions such as Airbnb and Uber is met by an equally strong pull from consumers across the world. The successful and proven SE models appear to be effective across many countries. Consumers from different countries appear to find accessing products and services in this way appealing. The differences between consumers from different geographic locations and cultures, who speak different languages and have different habits, do not appear to be a major challenge to the diffusion of SE models (Cheng & Zhang, 2019). They are becoming comfortable with social commerce and are moving from owning to temporary use (Puschman & Alt, 2016). In this way they find greater value both when offering and using products and services.

SE champions like Airbnb provide the online platform for the sharing. These platforms fulfil some functions, like processing the payment, but leave others, like describing the flat, to landlords that want to share. The person that wants to rent the room must have a sufficiently high trust and sufficiently low privacy concern to engage. While many organizations have built a trusting relationship with their consumers in traditional B2C e-commerce, it can still be a challenge for some organizations. Individual's privacy concerns are also a challenge. The individual must provide personal information to book the flat, but they are also vulnerable during their stay, in several ways including video surveillance. It is clear that the individual's privacy concerns are also elevated and compounded in SE because of the physical privacy risks that are added to information risks. Therefore, it is necessary to explore trust and privacy together. As the platform stops short of offering all the information and providing all the functionality, the landlord must build trust and reduce privacy concern.

Despite the role of the platform in bringing the two sides together, the renter will stay in a room of a stranger, not an organization with a recognized brand and reputation. There are higher risks and likelihood of distrust than a traditional hotel. Trust has been proven to play a role in online collaboration, particularly when value is exchanged. Trust plays a role in several online contexts such as sharing private information on online health services or paying with blockchain based currencies (Mendoza-Tello, Mora, Pujol-López, & Lytras, 2018). However, this role can change based on the context.

In addition to the usual risks of transacting online and unsecure private information, there is also the risk to physical privacy during the stay at the stranger's premises. Beyond their digital information, the renter can be vulnerable both optically and audibly during their stay. Privacy concerns are an important factor in the use of technology and sharing of information. As the technology and private information change in different contexts, the privacy concerns also change. Therefore, both trust and privacy models need to be adapted and extended to fully explain the environment of SE.

In addition to bringing the two sides together, the platform such as Airbnb, takes several steps to increase the limited trust such as making reviews from previous renters available. Nevertheless, the landlord must also reduce the feeling of risk, increase trust and reduce privacy concerns in the way they communicate the information about themselves and what they are offering. The landlord is an individual the renter has not met before and not an established organization with a recognized brand, so they only have a few words and pictures to achieve this.

Users in different countries have similar, if not the same, experiences on the Internet using popular global platforms. It is easy to neglect that different languages are still used, and they influence the interaction differently. The language used shapes the way a message is coded and decoded based on standardized language norms and culture. It has been proven that a language, like German or English, codes and decodes a message differently because of the differences in language norms and culture (Kohn & Hoffstaedter, 2017). Dimensions of culture such as power distance, collectivism versus individualism, femininity vs masculinity and uncertainty avoidance (Hofstede, Hofstede, & Minkov, 1997) have been found to influence how private individuals post content online (Pfeil, Zaphiris, & Ang, 2006). Nevertheless, the role of language and linguistics in information systems, and especially in the SE, has not been sufficiently covered. Therefore, the aim of this research is to explore what the trust building methods are and how privacy concerns are reduced in the SE, in German and English, using the case of Airbnb. Understanding how collaborative consumers build trust will also allow us to answer the question whether there are differences in building trust and reducing privacy concern in these two languages. Therefore, the research questions are:

- *Explore the trust building methods and how privacy concerns are reduced in the SE in German and English.*
- *Explore whether collaborative consumers build trust and reduce privacy concern in these two languages in the same way.*

Germany and England are chosen for several reasons: Firstly, Airbnb and other leaders in SE have been active in the two countries for several years now and are widely adopted. Secondly, Airbnb, Uber and other leaders in the platform and sharing economy offer their services across the two countries without significant differences. This means it is easier to attribute potential differences to the two languages. Thirdly, these are two large economies so the findings will be useful to a large audience directly. SE is expected to grow to 335 billion by 2025 (PWC, 2019) and the German and British economies include approximately 150 million people and are two of the six largest economies in the world.

The flat descriptions and profiles of the landlords offering their properties in Germany in German and in England in English are contrasted. To give the analysis validity the comparison between the two languages is made on the website of the same organization, Airbnb, and the profiles serve the same purpose. Furthermore, similar properties are chosen so that the influence of other demographic factors such as income and education are limited as far as possible. For this reason, descriptions of one-bedroom flats available are evaluated.

The findings indicate that language has a role in the interaction, but it is limited, and the platform norms and habits are more influential. Language plays a role primarily in three ways: Firstly, in the way the landlord expresses the benefits of the vacancy, secondly, the terms, conditions and fines and lastly the landlord's self-presentation in the personal profiles. The landlord does not usually reduce privacy concern but leaves this to the platform. There are several practical implications including clarifying the role of the landlord and the platform. This better understanding also informs the platform's decision to what degree it must adapt to different countries and for which parts of the business model this should be done.

## 2. Literature review

Participating in the SE requires trust between those collaborating (Bucher, Fieseler, & Lutz, 2016) and a sufficiently reduced privacy concern. A broad definition of trust is the willingness of one person to be vulnerable to the actions of someone else because of an expectation that the second person will take a specific action, without the ability to control or monitor the second person (Mayer, Davis, & Schoorman, 1995). A broad definition of the consumer's privacy concern is the concern over the improper access, collection, error and unauthorized secondary use of personal information (Smith, Milberg, & Burke, 1996). The literature review first covers the role of the platform and the landlord in building trust, then how privacy concerns are reduced and lastly the role of language in this context.

### 2.1. How the platform builds trust in the sharing economy

The platform performs some of the processes of a traditional retailer, but it is not the same. It is not a relationship between a seller and a buyer, but it is more like an organization with limited boundaries which creates trust-based relationships with partners outside the organization. The platform must match the landlord with the renter but may also moderate the exchange by keeping records and resolving conflicts. The renter must evaluate the trustworthiness of the platform (Ter Huurne, Ronteltap, Corten, & Buskens, 2017) and the landlord, so information that makes this decision easier is appreciated. Trust building can be divided into process-based, characteristic-based and institution-based (Meier, Lutkewitte, Mellewigt, & Decker, 2016).

#### 2.1.1. Trustworthiness of landlord

The platform supports trust in several ways by encouraging information sharing and aggregating data on those utilizing the platform. An example of how the platform encourages the sharing of relevant information is that landlords are encouraged to share personal strengths related to the flat being shared (Tussyadiah & Park, 2018). The data that is aggregated includes the previous transactions that are recorded. Information is collected to support the building of a reputation. When the platform keeps records to show the reputation of the participants it also encourages truth-telling behaviours and honesty. This increases the quality of information available, reduces search costs and makes it easier for someone new to sharing on this platform to gain the necessary familiarity with the process for them to share again (Möhlmann, 2015). Therefore, in addition to supporting the decision on whether to trust someone they encourage honesty which also supports trust. Similarly, these processes from the platform encourage a high quality of service which increases satisfaction and trust.

#### 2.1.2. Institutional trust

The two users considering transacting must also build trust within the parameters given by the platform. The platform can build trust is several ways. In addition to building reputational trust, the records kept can support action by the relevant institutions such as consumer protection agencies and law enforcement. For example, the platform presents the relevant regulation in a way that is easy for the renter to understand quickly. Therefore, the portal supports the building of trust but also the transference of trust between the landlord, the renter, the platform and other relevant institutions. Trust is transferred from one consumer to another, from the portal to the consumer (Lim, Sia, Lee, & Benbasat, 2006; Tams, 2012), and between institutions and the portal. The platform also supports trust in information disclosure by ensuring that the renter's information is protected, only used where necessary and only shared with whom is necessary.

Those engaging in the SE are in a common social network. In several contexts, social networks build the users trust (Hemmert, Kim, Kim, & Cho, 2016). Trust building in peer to peer networks is enhanced by peers' common needs, contributing their time, knowledge and information into the community.

### 2.2. How the landlord builds trust in the sharing economy

The landlord can build trust firstly with the flat description including pictures of the room and shared spaces, secondly their profile including a prominent picture of themselves (Ert, Fleischer, & Magen, 2016) and thirdly other listings of theirs. The landlord has an interest to maintain a high quality of service and to show their high quality so that they build trust. Trust in the landlord reduces the search cost for the renter as they can make a choice quicker by trusting the signals they are receiving. The renter must use the signals from the landlord to asses both the landlord and the accommodation. Those engaging in SE learn together, educating each other. The more similar the participants are, the more stable the trust building is over time (Cheng et al., 2016). The closer the landlord and renter come, the stronger the sense of commonality and characteristic-based trust (Zhang & Hamilton, 2010).

### 2.3. Reducing privacy concerns in the sharing economy

The SE introduces some information and physical privacy threats (Lutz, Hoffmann, Bucher, & Lutz, 2018). There are privacy concerns about how personal data is handled firstly by the platform, often referred to as institutional privacy, and secondly by the landlord. As the landlord has characteristics of a service provider and a lay user, they can be considered to cause both institutional and social privacy concerns. The platform plays a reduced role compared to a traditional B2C service provider and the landlord takes on processes that an institution would typically take, so the social media privacy concern is elevated. The renter is also sharing some personal information with other users in a similar way to participating in social media (Young & Quan-Haase, 2013). Physical privacy threats are caused by the consumer physically being present at the landlord's property. The consumer can be monitored in terms of their voice and their physical presence while they are there. Additionally, while the renter is using the flat, there are risks that their personal space will be violated. For example, the landlord can enter the flat when the renter does not want them to. After they check out, they leave behind fingerprints, hair etc. that can be used to extract their DNA code. Therefore, the SE brings an elevated social media concern along with physical privacy threats. The landlord also faces an elevated privacy threat, but this is not the focus of this research.

### 2.4. The language of building trust and reducing privacy concern

While platforms such as Airbnb play an important role for the exchange of value to be made, the landlord and renter need to cooperate. The cooperation involves risk for both sides in several ways including financial loss, security, safety and privacy. Information must be exchanged and a negotiation might also be needed, bringing some risk (Kong, Dirks, & Ferrin, 2014). The landlord starts this exchange with their profile and how they describe their property, so the language they use is important. Beyond some information conveyed explicitly, such as the size of the flat, there is other information implied through the language used. By conveying these implicit messages and seeing that they are received the negotiation moves forward, deadlocks are avoided (Belkin & Rothman, 2017), trust is built (Yao, Zhang, & Brett, 2017) and they move towards a result that satisfies both sides (Kong et al., 2014).

Language, beyond the explicit content, is the medium of the negotiation. The way each side communicate their message and how this message is understood is important. The message can be intricate with implied and subtle meaning. Creating and deciphering the message can be influenced by the language used and its emotiveness, grammar, norms, idioms and dialects. The language of the message can encapsulate and convey a culture that influences perception differently (Cheng & Zhang, 2019; Gumperz & Levinson, 1991). Differences in culture have been found to create different expectations between the landlord and the renter on Airbnb as different cultures bring different assumptions to the interaction (Cheng & Zhang, 2019). When someone deciphers a message in their language it can validate their culture and frame of reference (Forehand & Deshpandé, 2001; Koslow, Shamdasani, & Touchstone, 1994). Furthermore, the level of formality, and the language of distance and proximity can be different in the German and English language (Chiswick & Miller, 2005). Influences on trust and privacy are summarised in table 1.

Given that trust has both a psychological and social element and language influences both the individual's cognition of the message and wider society, it is useful to explore what methods of trust building and reducing privacy concern are employed in the context of Airbnb. As this platform offers many methods of communication with different purposes, such as presenting your property and yourself, it is also interesting to explore the specific purpose of each part of the property description and landlord profile.

### 3. Methodology

Data collection: The flat descriptions and personal profiles of landlords offering their properties in Germany in German and in England in English are contrasted. To increase the validity of the comparison several steps were taken in the data collection. Firstly, the comparison between the

| Table 1. How the platform influences trust and privacy concern | | |
|---|---|---|
| | **Platform ex. Airbnb** | **Examples from the literature** |
| Building trust | Insurance<br>Mediation<br>Intermediary with legal and regulatory institutions | (Puschman & Alt, 2016; Sutherland & Jarrahi, 2018)<br>(Sutherland & Jarrahi, 2018; Yang, Lee, Lee, & Koo, 2018)<br>(Cohen & Kietzmann, 2014; Sutherland & Jarrahi, 2018) |
| Reducing privacy concern | Privacy assurances<br>Provide security for personal and payment information | (Lutz et al., 2018; Ter Huurne et al., 2017)<br>(Puschman & Alt, 2016; Ter Huurne et al., 2017) |

two languages was made on the website of the same organization, Airbnb, and the descriptions served the same purpose. Furthermore, similar properties were chosen so that the influence of other demographic factors such as income and education are limited as far as possible. For this reason, descriptions offering a one-bedroom flat from 75 to 100 euro a night were included. To ensure that the sample is representative of each country, descriptions from several different regions were collected including both urban and suburban. No specific date was specified either, to avoid the possibility that the popularity of different periods in each country would cause discrepancies, such as different target audiences. As illustrated in Table 2, 800 property descriptions and 400 landlord profiles are included from both countries. The data was collected by the researchers by going through Airbnb flat listings and checking if they met the criteria that was outlined in this paragraph.

Data analysis: Qualitative analysis is used to find patterns in the language such as the structure, content and tone used. Open coding is used to find terms and phrases related to trust and privacy concerns. Some codes were identified from the literature, before the data was collected, but themes were also allowed to emerge from the data. This can be considered as directed content analysis because some initial theory from the literature gave some direction at the start of the coding process (Hsieh & Shannon, 2005). The codes identified from the literature were the level of formality, language of distance and proximity, implicit and implied meaning, emotiveness, grammar, norms, idioms and dialects, conveying a culture and assumptions. To increase the scientific rigor the data was coded by two researchers separately and they then discussed and agreed on shared codes.

The quantitative analysis compares the approaches to building trust and reducing privacy concern between Germany and England. The codes that were identified, are compared between the four groups listed in Table 2, to see if there are statistically significant differences in the regularity of their use. The data is binary as the property description either uses language to build trust and reduce privacy concern or it does not. Therefore, the Chi-Square test of independence is used (Bewick, Cheek, & Ball, 2004).

## 4. Results

The first three sections of the results cover how the landlord presents themselves, the positive aspects of the flat and the responsibilities of the renter. The fourth section identifies the main ways

| Table 2. Where the properties and landlords are located | | |
|---|---|---|
| | **Property description** | **Landlord profiles** |
| Germany, urban | 200 | 100 |
| Germany, suburban | 200 | 100 |
| England, urban | 200 | 100 |
| England, suburban | 200 | 100 |

language is used and the fifth section compares the use of language in the two countries quantitatively. Despite the data used being publicly available, specific excerpts are not included to avoid any personal privacy breaches.

### 4.1. Self-presentation in personal profiles

The language is positive and upbeat painting the picture of a happy person. Conveying happiness and avoiding anger builds trust (Belkin & Rothman, 2017). Often, the language used matched the purpose the flat would be rented for. Landlords with flats that would usually be used for leisure, due to their location near a popular tourist destination, promoted leisure and a relaxed lifestyle. The lifestyle of leisure is presented in the flat description and the landlord's profile where they often presented themselves as an embodiment of that lifestyle, showcasing the activities that are possible. For example, profiles of landlords in the German Alps showed how they enjoyed skiing. Similarly, landlords in the Peak District, an area popular for outdoor activities, presented themselves as people that enjoyed hiking and camping. The landlord's openness and friendliness, leaving their personal privacy vulnerable, may encourage reciprocal sharing of private information from the renter (Li, Sarathy, & Xu, 2010).

### 4.2. How the benefits are expressed in the flat descriptions

The analysis found that the benefits of the flat can be presented in a formal or informal way. The level of formality influences the level of emotiveness and how the language of distance and proximity is used. Several trust building features are utilized. Some trust building features are focused on reducing the perceived risk of certain aspects of the exchange. Firstly, the authenticity of some aspects of the flat is supported. Secondly, the reliability of the process of renting is also supported.

Despite the renter's privacy concern the landlord does not usually use their property description and profile to reduce privacy concern directly. This is despite some aspects of privacy, particularly physical privacy, being dependent on the landlord. Other aspects of privacy are also dependent on the landlord in some cases. The provision of Wi-Fi Internet access is common, but in some cases, the renter must use their personal account on devices such as connected audiovisual devices to access their content. The neglect of most landlords to engage on privacy concerns contrasts with the platform, Airbnb, that covers them with both legally worded information and more informal information. The landlord focuses on the immediate hedonic potential of the flat, the communal spirit and social reputation. These results are sumarised in figure 1.

### 4.3. Terms, conditions and fines in the flat descriptions

Beyond some information conveyed explicitly, such as the size of the flat, there is other information implied through the language used. The landlord can convey some boundaries and indicate how flexible they will be.

The landlord makes clear and explicit references to the contract with the renter. Terms, conditions and fines are often mentioned clearly. What are also conveyed, but not so explicitly, are the landlord boundaries and how strict they will be in applying the terms of the contract. The boundaries of acceptable behaviour from the guest, potential fines, and how a negotiation would happen are often covered. Some landlords try to keep a friendly informal tone while others convey a less informal, more professional, business tone for this part. Many landlords try to balance appearing inflexible and firm on the terms, while not appearing to be angry or aggressive. This is often achieved by conveying an inflexibility, by having more specific detail and language for the terms, without being aggressive. Another technique, is for the landlord to illustrate how professionally they approach this relationship, implying that they expect a similar level of professionalism. Often, the landlord will repeatedly use negative words such as 'no' and requests such as "please" close together. This part can be assertive, even passive-aggressive, but stops short of showing anger. There are very few cases where some aggression is conveyed. This aggression is usually showed in relation to the importance of the renter leaving on time. It is beneficial that most landlords manage to appear strict, without showing anger, as anger reduces trust (Belkin & Rothman, 2017).

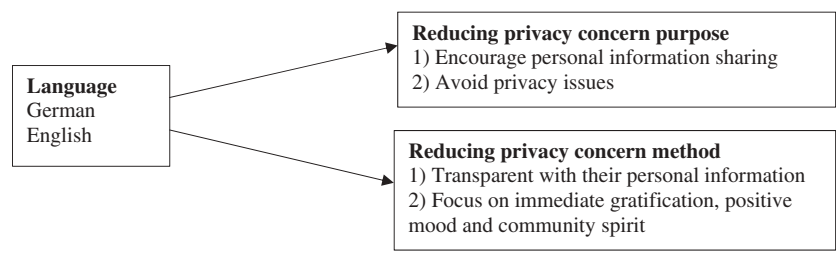

Figure 1. The role of language in reducing privacy concern in Airbnb.

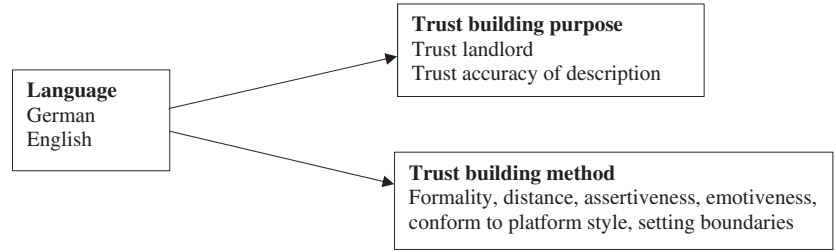

Figure 2. The role of language in building trust in Airbnb.

While Airbnb is usually perceived to facilitate sharing between consumers that are peers, there are also businesses with many employees and properties that utilize this platform. This influences how they communicate the terms of the contract. The businesses use more standardised language with legal terminology. These terms are formal and similar to what a hotel provides to protect its legal interests. The specific legal language used by businesses, shows their intention to utilize the relevant institutions if there is a problem and not just rely on the platform to resolve it. It can be concluded that private owners use language that conforms to the platform's social convention of having a friendly tone and rely on the platform to resolve problems, unlike businesses, that use the platform to access customers but do not entirely conform to the informal and friendly tone, and do not entirely rely on the platform to resolve conflicts.

### 4.4. The influence of the german and english language, similarities and differences

There is some limited influence of the German and English language on the message, such as differences in the message between urban and provincial flats, in both languages. The influence of the German and English language is mostly limited however. The platform, Airbnb, seems to have created a community with distinct social norms that brings a degree of common style, themes and tone, limiting the influence of German and English. The influence of the platform and community "modus operandi" is apparent. What may appear to be a social interaction between two individuals, is embedded within the platform's processes, regulations, social conventions and legal framework. Those sharing on Airbnb communicate with four primary purposes. The most extensive messages are from the landlord where they convey the terms and present themselves in the profile. There can be some limited negotiation and sharing of personal information. Figure 2 and table 3 summarise the results of the qualitative analysis.

### 4.5. Quantitative comparison of German and English flat descriptions

The quantitative analysis first compares the six approaches to building trust between Germany and England. The second stage compares two approaches to reducing privacy concern in the same way. As the data is binary, either the property description uses language in a way that builds trust and reduces privacy concern or it does not, the Chi-Square test of independence is used (Bewick et al., 2004).

| Table 3. How the platform and the landlord influence trust and privacy concern | | |
|---|---|---|
| | **Platform ex. Airbnb (based on literature)** | **Landlord (based on the data collected)** |
| Building trust | Insurance<br>Mediation<br>Intermediary with legal and regulatory institutions | Formal to informal<br>Distance to proximity<br>Assertive to passive aggressive, avoid anger<br>Emotiveness and humor<br>Conform to platform style<br>Setting boundaries |
| Reducing privacy concern | Privacy assurances<br>Provide security for personal information and payment details | Transparent with their personal information<br>Limited evidence of directly covering privacy and focus on:<br>• Immediate gratification<br>• Positive mood<br>• Community spirit and social reputation |

#### 4.5.1. Comparison of how German and English flat descriptions build trust

The first stage of the quantitative analysis compares the six approaches to building trust between Germany and England. Table 4 shows how regularly landlords use the six approaches in their language to build trust. Only the instances where the language style is used to achieve the desired effect are included. There are several methods to set boundaries beyond the way the language is used that are not included. An example where a boundary is set, without the style of language playing a role is the days the flat is available. The data is binary as the coding evaluates the whole description. For example, the whole description must be evaluated to conclude if it conforms to the platform style.

For the comparison of the six codes between the two countries the findings indicate that there is no difference in the use of language to build trust. This is in line with what the qualitative analysis suggests. As the Chi-Square statistic P-value, presented in table 5, is higher than 0.05 we accept the null hypothesis and conclude that the use of language is independent of the country. Therefore, it can be concluded that the two countries use language in the same way.

#### 4.5.2. Comparison of how German and English flat descriptions reduce privacy concern

The second stage of the quantitative analysis compares the two approaches to reducing privacy concern identified in the qualitative stage, between Germany and England. Tables 6 and 7 compare the two countries on transparency and avoiding privacy issues.

The findings of the comparison between the two countries indicate that there is a difference in the first approach but not in the second. There is a statistically significant difference in the landlord's transparency with personal information. Landlords in Germany are less transparent

| Table 4. The number of property descriptions that use this method of building trust with language | | | | | | |
|---|---|---|---|---|---|---|
| | Formal to informal | Distance to proximity | Assertive to passive aggressive | Emotiveness and humor | Conform to platform style | Setting boundaries |
| Germany | 70 | 89 | 21 | 78 | 221 | 151 |
| England | 88 | 70 | 34 | 59 | 242 | 128 |

| Table 5. Comparison of how the landlord uses language to build trust in Germany and England | | |
|---|---|---|
| **Comparison** | **P-value** | **N** |
| 1. Formal to informal: Germany-England | 0.110 | 800 |
| 2. Distance to proximity: Germany-England | 0.092 | 800 |
| 3. Assertive to passive aggressive: Germany-England | 0.069 | 800 |
| 4. Emotiveness and humor: Germany-England | 0.075 | 800 |
| 5. Conform to platform style: Germany-England | 0.133 | 800 |
| 6. Setting boundaries: Germany-England | 0.088 | 800 |

Degrees of freedom = 1, significance level = 0.05, N = total sample size of both samples being compared

| Table 6. The number of property descriptions that use this method of reducing privacy concern with language | | |
|---|---|---|
| | **Transparent with their personal information** | **Avoid privacy issues** |
| Germany | 44 | 388 |
| England | 73 | 392 |

| Table 7. Comparison of how the landlord uses language to reduce privacy concern in Germany and England | | |
|---|---|---|
| **Comparison** | **P-value** | **N** |
| 1. Transparent with their personal information: Germany-England | 0.004 | 800 |
| 2. Avoid privacy issues: Germany-England | 0.365 | 800 |

Degrees of freedom = 1, significance level = 0.05, N = total sample size of both samples being compared with their personal information than those in England. This is the only statistically significant difference in the use of language identified. Lastly, the results suggest landlords in both countries avoid privacy issues.

## 5. Discussion

The language used in a message encapsulates a culture with certain implied norms and encourages those receiving the message to follow those norms (Cheng & Zhang, 2019; Kohn & Hoffstaedter, 2017). As there are limited differences between German and English speakers, this indicates that they have been linguistically assimilated by the platform. As people spend more time on the Internet using a small number of platforms and less time face to face in their communities, native speakers and their implied culture have a loosening grip on language. As the influence of the platforms increases more importance is given to following the platform convention and its implied norms. The breadth of language used may also be reduced because of the strength of the platforms influence. Language communicates a culture but also influences it (Gumperz & Levinson, 1991) so this shared language may also lead to a shared culture. While the local language may have a reduced influence, using Airbnb in Germany, with the German language may legitimize the platform in this country.

### 5.1. Theoretical implications

The initial contribution to theory of this research is bringing the literature of information systems, the SE, trust, privacy and linguistics together. Furthermore, the division of responsibility for building trust and reducing privacy concern between the platform and the landlord are clarified. The landlord focuses on building trust through their property description and profile while the platform supports trust and decreases privacy concerns.

The third theoretic contribution is identifying the linguistic approaches to building trust and reducing privacy concern. For building trust these are: (1) the level of formality, (2) distance and proximity, (3) emotiveness and humor, (4) being assertive, including passive aggressive, but avoiding anger, (5) conformity to the platform language style and terminology and lastly (6) setting boundaries. Despite the elevated and compounded privacy concern, the landlord does not usually use their property description and profile to reduce privacy concern directly. Most avoid privacy issues, and some are transparent with their own personal information to encourage reciprocity. This is in contrast to the effort the landlord makes to build trust. Both the platform and the landlord build trust with the language they use. It appears that the landlord delegates reducing privacy concern to the platform Airbnb. While this may be suitable for personal information, such as banking details, that are not shared with the landlord, this is less suitable for physical privacy that is shared with the landlord. There is a void of neither the platform or the landlord, covering physical privacy. This may not necessarily reduce the willingness of the renter to use a flat for two reasons: Firstly, there are many cases where there is a privacy concern, but an online service is nevertheless used (Young & Quan-Haase, 2013). Secondly, there is evidence that if a user does not have sufficient information to make a decision on privacy, they may use the service regardless or relinquish their decision to a proxy (Dinev, Albano, Xu, D'Atri, & Hart, 2015), in this case the platform. Thirdly, it has been shown that social reputation is an effective motivator to participate in online value exchange (de Rivera, Gordo, Cassidy, & Apesteguia, 2016). Fourthly, online information that creates a positive mood encourages information sharing (Wakefield, 2013). Fifth, as the landlord is providing relevant information and may also have their own physical privacy compromised, the renter may consider this a fair information exchange according to privacy calculus and social contract theory (Li et al., 2010).

### 5.2. Practical implications

Firstly, there appears to be limited benefit in adapting the platform for Germany and England because of the limited role of language compared to the platform norm in communication. This is partly contrary to research comparing the influence of Western and Chinese culture on Airbnb interactions, where more significant differences were found (Cheng & Zhang, 2019). It appears that the efforts of the platform to reduce risk in several ways, including the aggregated information they collect, are effective enough to make the users rely on them and embed their communication within them. Airbnb appears to have achieved a suitable balance between the conflicting interests of the two parties.

In terms of building trust there are practical implications for the landlord and the platform. The constituent participants of the SE should be clear about where the other participant is building trust effectively and where they need to build trust. An example is that the landlord should build trust and reduce privacy concern for physical privacy, while privacy of financial information related to the payment can be covered by the platform.

It is important for platforms, landlords and the related institutions to support trust transference. The literature identified that the portal supports the building of trust, but also the transference of trust. Trust is transferred from one consumer to another, but also from the portal to the consumer (Lim et al., 2006; Tams, 2012). Those presenting their service for sharing may follow the platform and community norm to enhance the transference of trust. When the landlord aligns their message to that of Airbnb, they not only show evidence that they are part of a community, but they also show that they are part of Airbnb, without being directly employed. It has been shown by Airbnb and Uber that the boundaries of an organization can be less strict than what was traditionally the case before the SE business model. Therefore, the landlord, acting more like a part time employee of Airbnb adopts the corporate language and communication style. During this exchange the three parties involved, the renter, the landlord and the platform have similar, but not necessarily identical interests. By aligning the language, the landlord is also indicating that the interests appear to be as aligned as possible. One example of this is that flats that are usually used for a more relaxed, slow paced recreational activity, have language in the landlord's description of the flat that reflects this.

## 6. Conclusions

This research compared the landlord property posts and their personal profiles on Airbnb in Germany and England. Qualitative analysis was used to explore how the landlord used their language to build trust and reduce privacy concern. The landlord builds trust both in their flat description and their personal profile. In their profiles they encourage private information sharing by sharing their own private information but beyond that they usually relinquish reducing privacy concern to the platform. Physical privacy is not addressed by the landlord or the platform.

This research found that the differences in presentation on Airbnb in German and English are limited. The platform appears to be a bigger influence on how the landlord presents their property and themselves. We can say that the platform has choreographed this interaction between the participants sufficiently, so that different languages have a limited influence on the outcome. There seems to be a limited benefit to adapting the platform for different countries.

A limitation of this research is that it evaluates the German and English words used but not how they are understood. Given the hundreds of posts analyzed, what is used by the majority is a strong indication of what is effective. Nevertheless, it is useful to distinguish between the words used (Lexical) and the meaning (conceptual) (Luna & Peracchio, 2001). It would therefore be useful for future research to explore how the German and English posts are understood. Furthermore, this research only contrasted two European countries, so it would be useful for future research to cover more countries and languages.

**Funding**
The authors received no direct funding for this research.

**Author details**
Alex Zarifis[1]
E-mail: a.zarifis@lboro.ac.uk
Richard Ingham[2]
E-mail: ringham04@yahoo.co.uk
Julia Kroenung[3]
E-mail: kroenung@bwl.uni-mannheim.de
[1] School of Business and Economics, Lougborough University, Loughborough, UK.
[2] English, Linguistics and Cultural Studies, University of Westminster, London, UK.
[3] Business School, University of Mannheim, Mannheim, Germany.

**Cover Image**
Source: Author.

*Cogent Business & Management* (ISSN: 2331-1975) is published by Cogent OA, part of Taylor & Francis Group.

**Publishing with Cogent OA ensures:**

- Immediate, universal access to your article on publication
- High visibility and discoverability via the Cogent OA website as well as Taylor & Francis Online
- Download and citation statistics for your article
- Rapid online publication
- Input from, and dialog with, expert editors and editorial boards
- Retention of full copyright of your article
- Guaranteed legacy preservation of your article
- Discounts and waivers for authors in developing regions

Submit your manuscript to a Cogent OA journal at www.CogentOA.com

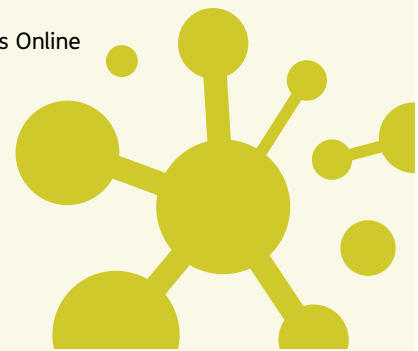